\documentstyle[prl,aps,epsfig]{revtex}
\def\func#1{\mathop{\rm #1{}}\nolimits}
\def\eps{\varepsilon}
\begin{document}
\draft
\twocolumn[\hsize\textwidth\columnwidth\hsize\csname@twocolumnfalse\endcsname
\title{Correlations in the Adiabatic Response of Chaotic Systems}
\author{Ophir M. Auslaender\cite{byline1,byline2} and Shmuel Fishman\\
Department of Physics, Technion, Haifa 32000, Israel}

\maketitle
\begin{abstract}
Adiabatic variation of the parameters of a chaotic system results
in a fluctuating reaction force. In the leading order in the
adiabaticity parameter, a dissipative force, that is present in
classical mechanics was found to vanish in quantum mechanics. On
the time scale $t$, this force is proportional to $I(t)$, the
integral of the force-force correlation function over time $t$. In
order to understand the crossover between the classical and the
quantum mechanical behavior we calculated $I(t)$ in random matrix
theory. We found that for systems belonging to the Gaussian
unitary ensemble this crossover takes place at a characteristic
time (proportional to the Heisenberg time) and for longer times
$I(t)$ practically vanishes, resulting in vanishing dissipation.
For systems belonging to the Gaussian orthogonal ensemble
$I(t)\sim 1/t$ and there is no such characteristic time. $I(t)$ is
calculated for various models and the relation to experiment is
discussed.
\end{abstract}
]

Friction is usually described in statistical physics as transfer
of energy to a system that consists of an infinite number of
degrees of freedom in an irreversible way. The irreversibility is
a result of the complexity of the motion of the many body system.
A natural question is whether the coupling to a small chaotic
system can exhibit dissipation resulting from its complex
dynamics. Mechanisms of this nature were first introduced in the
context of nuclear physics \cite{Hill52}, and various aspects of
it were discussed in the past
\cite{Blocki78,Wilkinson95,Austin92,Berry93,Jarzynski95}. In the
present work we study, following Berry and Robbins (BR)
\cite{Berry93} and Jarzynski \cite{Jarzynski95}, a specific model
where the chaotic system is defined by the Hamiltonian ${\cal
H}\left({\bf R},{\bf z}\right)$ where ${\bf z}\equiv\left({\bf
p},{\bf r}\right)$ are the phase space coordinates of the chaotic
system and ${\bf R}$ is an adiabatically varying parameter. The
system is chaotic for each value of ${\bf R}$. The crucial feature
of the system we studied in this work \cite{we} is that it
exhibits a wide separation of time scales - the evolution of the
chaotic system, characterized by the time scale $T_{\mbox{fast}}$,
is so rapid that it explores all of the phase space accessible to
it before the parameter ${\bf R}$, characterized by the time scale
$T_{\mbox{slow}}$, changes appreciably. The adiabaticity parameter
is $\eps\sim T_{\mbox{fast}}/ T_{\mbox{slow}}$. The average
generalized force, that is applied by the chaotic system in
question on the system that forces the adiabatic variation of
${\bf R}$, is given by:
\begin{equation}\label{AvgForce}
{\bf F}(\tau_a)=-\int {\bf dz}~\rho\left({\bf z},\tau_a \right)
\partial_{\bf R}{{\cal H}\left({\bf z},{\bf R}(\tau_a)\right)},
\end{equation}
where $\rho({\bf z},\tau_a)$ is a normalized probability density
in the fast particle phase space. If ${\bf R}$ is a position space
coordinate, ${\bf F}(\tau_a)$ reduces to a regular force. It can
be expanded in powers of the adiabaticity parameter $\eps$ as was
done by BR \cite{Berry93} and Jarzynski \cite{Jarzynski95}. The
present work follows closely the formalism of BR.

Berry and Robbins \cite{Berry93} were able to calculate the force acting
on the slow particle up to first order in $\eps$:
\begin{equation}\label{BRforce}
{\bf F}\approx {\bf F}_0+\eps{\bf F}_1.
\end{equation}
To leading order, the force is given by the classical analogue of
the Born-Oppenheimer force
${F_0}_i(\tau_a)=-\partial_{R_i}{E\left({\bf R}\right)}$ where
$E\left({\bf R}\right)$ is chosen such that the phase space volume
of the fast particle, $\Omega\left(E\left({\bf R}\right),{\bf
R}\right)$, is constant \cite{Ott79}. The leading correction to
${\bf F}_0$ includes a velocity dependent force ${\bf
F}_1(\tau_a)$, and describes two qualitatively different forces.
The first of these is geometric magnetism, that is related to the
Berry phase \cite{bphase}. The second one is related to
deterministic friction \cite{Wilkinson95,Jarzynski92,Berry93}. A
central question is under which conditions friction due to the
velocity dependent force ${\bf F}_1$ is found.

For the exploration of the existence of friction it is sufficient
to study the case where ${\bf R}$ is replaced by a scalar $X$. In
this case the velocity dependent force ${\bf F_1}(\tau_a)$ reduces
to
\begin{equation}\label{F1}
F_1(\tau_a)=-\dot{X}~ \Sigma^{-1}\partial_{E}\left[\Sigma~
I(t=\infty)\right]_{E=E(X)},
\end{equation}
where $\Sigma(E,X)\equiv\partial_{E}\Omega(E,X)$ and:
\begin{equation}\label{InDef}
I(t)=\int_0^t C(t')dt'.
\end{equation}
The fluctuating force-force correlation function at time difference $t'$ is:
\begin{equation}\label{BRCij}
C(t')\equiv\Big< \left(\partial_X \tilde{\cal H} \right)_{t'}
\left(\partial_X \tilde{\cal H}\right)_0\Big>_{E,X},
\end{equation}
where $\left<\ldots\right>_{E,X}$ denotes the microcanonical
average over the energy surface with a fixed value of the
parameter $X$. The volume enclosed by this surface is an adiabatic
invariant \cite{Ott79}. The fluctuation of the energy is
$\tilde{\cal H}\left({\bf z}, X \right)\equiv {\cal H}\left({\bf
z},X \right)-E(X)$. A finite value of the integral $I(\infty)$ is
required for friction. Within classical mechanics this integral is
indeed positive. It was shown by BR that in quantum mechanics the
dissipative part of the force ${\bf F}_1$ vanishes. To understand
the reason for this discordance (\ref{BRCij}) was calculated and
in quantum mechanics it takes the form:
\begin{equation}\label{BRCijQ}
C(t)=\sum_{m \neq n}~ \left|\left<n\left|\partial_X{\widehat {\cal
H}}\right|m\right> \right|^2
\cos\left[\frac{t}{\hbar}\left(E_n-E_m\right)\right],
\end{equation}
where the $E_m$ are the eigenenergies of the chaotic system. The
initial state is $n$ and the dependence on it is not important for
the present work. The sum is over $m$. For this correlation
function the integral $I(\infty)$ of (\ref{InDef}) vanishes. In
order to understand how the crossover between the classical and
the quantum behavior occurs, it is instructive to calculate the
integral of the correlation function over a finite time. Taking
the classical limit $\hbar\rightarrow 0$ for any finite $t$ and
then the limit $t\rightarrow\infty$ should result in a
non-vanishing value of $I(\infty)$, while for any finite value of
$\hbar$, $I(\infty)$ should vanish. The friction on the time scale
$t$ is proportional to $I(t)$ as can easily be inferred from
(\ref{F1}). For systems whose classical dynamics is chaotic, the
energy levels are distributed according to random matrix theory
(RMT) \cite{Bohigas84}. The long time behavior of $I(t)$ is
determined by the levels nearest to $n$, namely $n\pm1$, as can be
seen from (\ref{BRCijQ}). The natural question to ask is whether
there is a characteristic time scale for the crossover between the
quantum behavior of the integral $I(t)$ and its classical
behavior. The most na\"{\i}ve answer to this question is that the
characteristic time scale is the Heisenberg time. On the other
hand, one can argue that there is no time scale for this crossover
at all \cite{izrailev}. In RMT the probability for two consecutive
levels to be separated by an energy difference $s$ behaves like
$s^\beta$ for small spacings \cite{Porter65,Mehta91}.
Consequently, $\left<I_\beta(t)\right>\sim t^{-\beta}$ for long
times. Here $\left<\ldots\right>$ denotes the RMT ensemble
average, and $I_\beta$ is the integral (\ref{InDef}) for some
$\beta$. The $\left<\ldots\right>$ will be dropped from $C_\beta$
and $I_\beta$ for notational simplicity.

For the Gaussian orthogonal ensemble (GOE) ($\beta=1$) one indeed
finds that $I_\beta(t)$ decays like $1/t$, but for the Gaussian
unitary ensemble (GUE) ($\beta=2$) one finds that it decays like a
Gaussian with a characteristic time proportional to the Heisenberg
time or vanishes after the Heisenberg time depending on the
parametric dependence on $X$. Why is the nature of the decay of
$I_\beta(t)$ important? There is the quantum-classical discordance
that has already been mentioned, and one would like to analyze the
scale that is required to observe the crossover between the
regimes. In addition, the model discussed in the present work is
relevant for some experimental situations. Consider for example a
molecular beam prepared in a classical configuration, where
initially many levels are substantially populated. The beam
travels in a slowly varying field \cite{DCPC}. Consequently the
internal dynamics in the molecules is in a slowly varying
potential. Another example is of quantum dots where parameters are
varied adiabatically like in pumping experiments, but with closed
dots, so that their spectrum is discrete \cite{marcus}. A dramatic
change in the energy absorption is predicted when a magnetic field
is applied and the transition from GOE to GUE takes place.

The time over which the correlation function decays should be
compared with other time scales present in the specific system
studied. One such time scale is $T_2\sim\eps^{-2}$, which is the
time scale for the breakdown of the first order of the multiple
scale expansion in $\eps$. Non-perturbative effects, such as
Landau-Zener tunneling, become important on a time scale of
$T_{LZ}$. In realistic experiments there is also the time scale
for quantum decoherence $T_\phi$. In order to observe the
classical to quantum crossover discussed in the present work
$\left<I_\beta(t)\right>$ should exhibit substantial decay for
$t\ll \min\left(T_2,T_{LZ}\right)$ and of the order of  $T_\phi$.
The energy absorption by the internal degrees of freedom is
proportional to $I_\beta(T_\phi)$.

\begin{figure}
\centering\epsfig{file=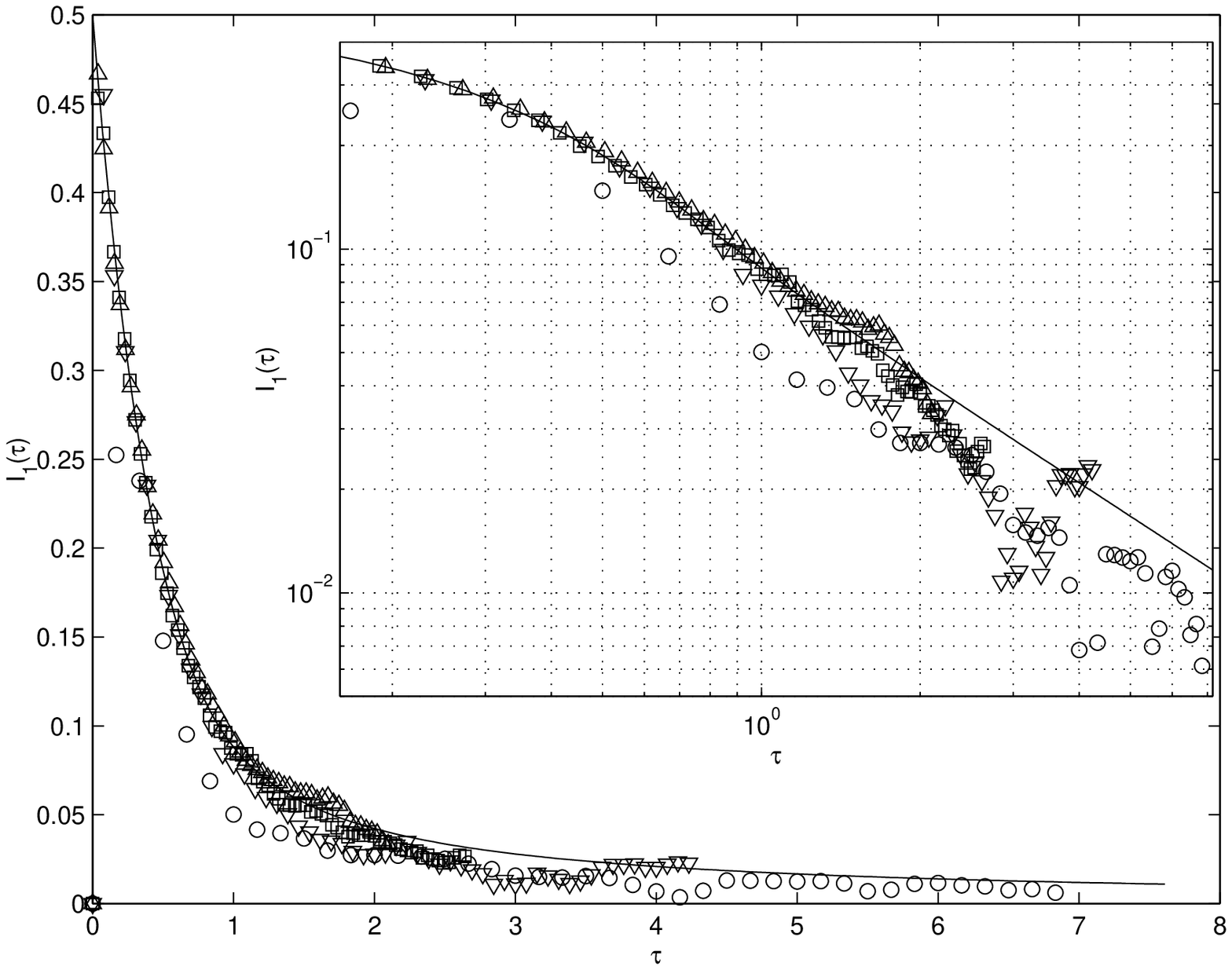,width=6.5cm,angle=90} \caption{The
integral of the correlation function for GOE. Numerical results
for $N=3$ (${\scriptscriptstyle\bigcirc}$), $N=13$
($\scriptstyle\bigtriangledown$), $N=53$ ($\scriptstyle\Box$) and
$N=103$ ($\scriptstyle\bigtriangleup$) are shown. Also shown is
the large N approximation (Eq.~\protect\ref{IGOE}) (line).}
\label{figGOE}
\end{figure}

\begin{figure}
\centering\epsfig{file=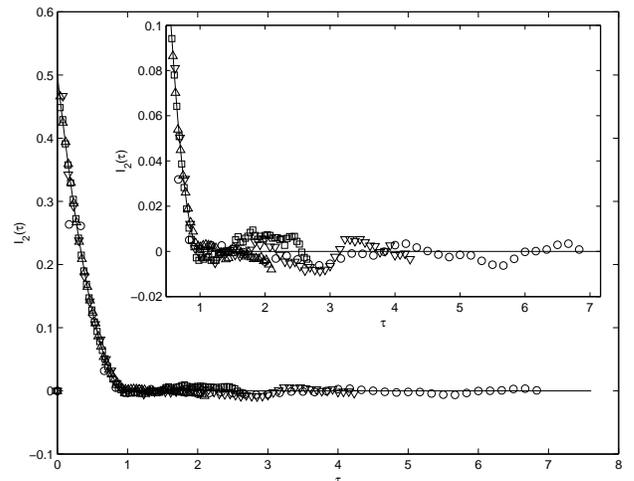,width=6.5cm,angle=90}
\caption{Similar to Fig. \ref{figGOE} for GUE, compared to the
large N approximation (Eq.~\protect\ref{IGUE}).} \label{figGUE}
\end{figure}

We shall use the Hamiltonian introduced by Austin and Wilkinson
\cite{Austin92} and model a parameter dependent system by the
$N\times N$ random matrix:
\begin{equation}\label{model}
H(X)=H_1 \cos{X}+H_2\sin{X},
\end{equation}
where $H_{1,2}$ are $N\times N$ random matrices from the same RMT
ensemble. There are three advantages to working with $H(X)$: (a)
it belongs to the same ensemble that $H_{1,2}$ belong to; (b) the
derivatives of its matrix elements belong to the same ensemble;
(c) the matrices $H(X)$ and $H'(X)\equiv dH(X)/dX$ are
statistically independent. If we insert $H(X)$ and $H'(X)$ into
Eq.~\ref{BRCijQ}, and then perform the ensemble average, we
obtain:
\begin{equation}\label{CnDef}
C_\beta(t)=\Bigg<\sum_{m\neq
n}\left|H'(X)_{n,m}\right|^2\cos\left[\frac{t}{\hbar}
\left(E_n-E_m\right)\right]\Bigg>,
\end{equation}
where $H'(X)_{n,m} \equiv \left<n\left|d{\widehat
H}(X)/dX\right|m\right>$ and $\left<\ldots\right>$ denotes RMT
ensemble averaging.

The statistical independence of $H'(X)$ and $H(X)$ implies
\begin{equation}\label{CnNoCorr}
C_\beta(t)=\beta\mu^2\sum_{m\neq n}
\Bigg<\cos\left[\frac{t}{\hbar}\left(E_n-E_m\right)\right]\Bigg>.
\end{equation}
The fact that $H'(X)$ belongs to the same ensemble as $H(X)$,
leading to $\big<\left|H'(X)_{n,m}\right|^2 \big>= \big<
\left|H_{n,m}\right|^2 \big>=\beta \mu^2$ for $m\ne n$, was used.

We would like to make the connection between
$C_\beta(t)/\beta\mu^2$ and the form factor:
\begin{equation}\label{formfactor}
K(t)=\int\left[\frac{1}{\overline{\rho}^2(E)}
\Big<\rho(E+\epsilon/2\overline{\rho}) \rho(E-\epsilon/2\overline{\rho})\Big>-
1\right]e^{i2\pi\epsilon \tau}d\epsilon,
\end{equation}
where $\rho(E)=\sum_i \delta(E_i-E)$ is the density of states and
$\overline{\rho}(E)$ is the smoothed density of states. The
variable $\epsilon$ is the energy measured in units of the mean
level spacing $1/\overline{\rho}(E)$ and $\tau=t/T_H$ is time in
units of the Heisenberg time, $T_H=h\overline{\rho}(E)$. In what
follows units where $\beta\mu^2T_H=1$ will be used.
Eq.~\ref{CnNoCorr} can be written in the following form:
\begin{eqnarray}
\frac{C_\beta(\tau)}{\beta \mu^2} &=&
\int\left[\frac{1}{\overline{\rho}^2(E)}
\Big<\rho(E+\epsilon/2\overline{\rho})\rho(E-\epsilon/2\overline{\rho})\Big>
-\delta(\epsilon) \right]\nonumber \\ &&\times
e^{i2\pi\eps\tau}d\epsilon.
\end{eqnarray}
Comparing the last equation with
(\ref{formfactor}) one can see that:
\begin{equation}
C_\beta(\tau)/\beta \mu^2=K(\tau)+\delta(\tau)-1.
\end{equation}
In this work we are mainly interested in the time integral of the
correlation function (\ref{InDef}):
\begin{equation}\label{IandK}
  I_\beta(\tau)= 1/2-\int_0^{\tau}d\tau'
\Big(1-K(\tau')\Big).
\end{equation}
In the limit $\tau \rightarrow \infty$ the RHS is just
$R_2(\epsilon=0)$, the two point spectral correlation function at
zero energy separation. It vanishes as a result level repulsion.

In order to perform actual calculations we make use of the well
known form factor for GOE and GUE \cite{Mehta91}. It is standard
to define $b(\tau)=1-K(\tau).$ For GOE (see \cite{Mehta91}
p.~137): $b(\tau)=1-2\tau+\tau\ln b_+$ for $\tau\leq1$ and
$b(\tau)=-1+\tau\ln [ b_+/b_-]$ for $\tau\geq1$ where $b_\pm=
2\tau \pm 1$, leading to:
\begin{equation}\label{IGOE}
  I_1(\tau)=\left\{
  \begin{array}{lr}
\frac12-\left[\frac54\left(\tau-\tau^2\right)+\frac12
\left(\tau^2-\frac14\right)\ln\left[1+2\tau\right]\right] &
\tau\leq1 \\
\frac12-\left[\frac12\left(1-\tau\right)+\frac12\left(\tau^2-\frac14\right)\ln
\left[\frac{2\tau+1}{2\tau-1}\right]\right]& \tau\geq1
  \end{array}
  \right..
\end{equation}
For $\tau\rightarrow\infty$ it falls off asymptotically as
\begin{equation} \label{IGOEa}
I_1(\tau) \sim 1/12\tau.
\end{equation}
For GUE (see \cite{Mehta91} p.~95): $b(\tau)=1-\tau$ for
$\tau\leq1$ and it vanishes for $\tau\geq1$, from which one
obtains:
\begin{equation}\label{IGUE}
  I_2(\tau)=\left\{
  \begin{array}{cr}
    1/2-\left[\tau-\tau^2/2\right] & \tau\leq1 \\
    0 & \tau\geq1
  \end{array}
  \right..
\end{equation}
The behavior for the Gaussian symplectic ensemble (GSE) is very
similar.

In order to compare the analytical results that hold in the
infinite $N$ limit with numerical data, ensembles of $N\times N$
matrices $H_1$ and $H_2$ of (\ref{model}), belonging to GOE or GUE
were generated numerically and the results are presented in
Figs.~\ref{figGOE}~\&~\ref{figGUE}.

The model (\ref{model}) is very specific in its dependence on the
parameter $X$. An important property of this model is the
statistical independence between $H(X)$ and $H'(X)$. Such
independence holds to a good approximation for disordered systems
\cite{lerner}. It is reasonable to make this approximation also
for RMT models of chaotic systems. The reason is that most
eigenstates look random, are statistically independent of the
eigenvalues and therefore for many types of perturbations the
matrix elements of $H'(X)$ will look random and independent of the
spectrum. Although this argument is reasonable for many types of
parametric dependence it is clearly not general. For the
asymptotic behavior much less is required, since the long time
asymptotics is dominated by the nearest neighbor level spacing.
The reason for this dominance is that if $\tau\gg 1$ the terms in
the sum (\ref{CnDef}) oscillate wildly as a function of $m$, so
that the important net contribution is from the terms nearest to
being stationary. These are obviously $m=n\pm1$. Introducing also
the {\em crucial} assumption that that fluctuations of
$\left|\left(H'(X)\right)_{n,n \pm 1}\right|^2$ can be ignored, we
find that in RMT:
\begin{equation}\label{CnApproxLong2}
C_\beta(\tau) /\beta\mu^2 \approx 2 \int_0^\infty \!\!\!\!ds~
P_\beta(s)\cos{(2 \pi \tau s)},
\end{equation}
where $s$ is the nearest neighbor level spacing in units of the
mean level spacing $\Delta E=1/\overline{\rho}(0)$ and
$P_\beta(s)$ is the distribution of $s$. The integral of the
correlation function is:
\begin{equation}\label{InApproxLong}
I_\beta(\tau) \approx
\frac{1}{\pi}\int_0^\infty \!\!\!\!ds~
\frac{P_\beta(s)}{s}\sin(2 \pi\tau s).
\end{equation}
For the nearest neighbor level spacing distribution we use the
Wigner surmise (see Eq.~202 in \cite{Porter65}):
\begin{equation}\label{Psbeta}
P_\beta(s)=c~s^\beta \exp{\left[-a s^2\right]}
\end{equation}
with $a=\pi/4$ and $c=\pi/2$ for GOE, and $a=4/\pi$ and
$c=32/\pi^2$ for GUE. The integral (\ref{InApproxLong}) can be
calculated for these distributions. For GOE one finds that for
large $\tau$ it takes the form: $I_{1}(\tau) \sim 1/4\pi\tau$,
that is extremely close to (\ref{IGOEa}). For GUE one finds:
$I_{2}(\tau)=2\tau \exp{\left[-\pi^3 \tau^2/4\right]}$, that for
large $\tau$ is very close to (\ref{IGUE}). Here too the behavior
for GSE is very similar to that of GUE.

We found that for the RMT ensembles $I_\beta(\tau)$ of
(\ref{InDef}) is dominated by the nearest neighbor level spacings
resulting in the approximation (\ref{InApproxLong}) for
$I_\beta(\tau)$. Under the assumptions leading to this
approximation $I_\beta(\tau)$ can be calculated for the nearest
neighbor level spacing distribution (\ref{Psbeta}) with arbitrary
$\beta$, $a$ and $c$ even if $H$ does not belong to an invariant
Gaussian RMT ensemble. The integral of the correlation function
(\ref{InApproxLong}) takes the form $I_\beta(\tau)= \left(c/\pi a
^{\beta/2}\right){\cal I}_\beta (y)$ where ${\cal
I}_\beta(y)\equiv\int_0^\infty \!\!\!\!ds~ s^{\beta-1}
  \exp{[-s^2]}\sin{s y}$ with $y=2 \pi \tau /\sqrt{a}$.
It satisfies the ordinary ordinary differential equation:
\begin{equation}\label{DiffEqn}
   \frac{d^2}{dy^2}{{\cal I}}_\beta(y)+\frac{y}{2}\frac{d}{dy}
  {{\cal I}}_\beta(y)+\frac{\beta}2
  {\cal I}_\beta(y)=0
\end{equation}
with the boundary conditions ${\cal I}_\beta(0)=0$ and $d{\cal
I_\beta}(0)/dy=\Gamma\left[(\beta+1)/2\right]/2$. Making the
substitution ${\cal
I}_\beta(y)=f_\beta(y/\sqrt{2})\exp{\left[-y^2/8\right]}$ and
changing to the variable $x=y/\sqrt{2}$, one arrives at a new
differential equation for $f$ that is a well known equation, and
its solutions are Parabolic Cylinder Functions \cite{AS}:
$\func{U}\left[1/2 -\beta,x\right], \func{V}\left[1/2
-\beta,x\right]$. For arbitrary $\beta$, its solution is a linear
combination of these two functions. From the initial conditions
and the known behavior of $U$ and $V$ at $0$ and $\infty$ the
asymptotic behavior of $I_\beta(\tau)$ is found to be of two
types. For $\beta\neq2n$ ($n=1,2,3,\ldots$): $I_\beta(\tau)\sim
\tau^{-\beta}$, as $\tau\rightarrow\infty$, while for $\beta=2n$
($n=1,2,3,\ldots$):
$I_\beta(\tau)\sim\exp{\left[-\pi^2\tau^2/a\right]}\tau^{\beta-1}$,
as $\tau\rightarrow\infty$. This is precisely the type of behavior
found for the RMT ensembles treated explicitly.

It was shown that for the model (\ref{model}) there is a big
qualitative difference between the integrals of the correlation
functions of the RMT ensembles (Eqs.~\ref{IGOE}~\&~\ref{IGUE}). It
was argued that the qualitative difference holds for a wide
variety of models. For models where the mean level spacing is
(\ref{Psbeta}), $I_\beta(\tau)$ decays like the power-law
$\tau^{-\beta}$, except for $\beta$ that are positive even
integers, for which it decays like a Gaussian with a
characteristic time that is proportional to the Heisenberg time.
Only for even values of $\beta$ the integral (\ref{InApproxLong})
can be written as an integral over an analytic function in the
interval $[-\infty,\infty]$, and is dominated by a saddle point in
the complex plane.

We have benefited from discussions with M. Berry, E. Bogomolny, D.
Cohen, B. Eckhardt, J. Feinberg, C. Jarzynski, C. Marcus, E. Ott,
R. Prange, J. Robbins, H.-J. St\"{o}ckmann, M. Wilkinson and M.
Zirnbauer. We thank in particular B. Simons for extremely
illuminating remarks. This research was supported in part by the
U.S.--Israel Binational Science Foundation (BSF), by the Minerva
Center for Non-linear Physics of Complex Systems, by the Israel
Science Foundation, by the Niedersachsen Ministry of Science
(Germany) and by the Fund for Promotion of Research at the
Technion. One of us (SF) would like to thank R. Prange for the
hospitality at the University of Maryland where this work was
completed.

\end{document}